\documentclass[12pt]{article}
\usepackage[utf8]{inputenc}
\usepackage[british]{babel}
\usepackage{cmap}
\usepackage{lmodern}

\usepackage{amssymb, amsmath, amsthm}
\usepackage[a4paper,top=25mm,bottom=25mm,left=25mm,right=25mm]{geometry}
\usepackage{ragged2e}

\usepackage{authblk} % for headings
\usepackage{pifont}
\usepackage{graphicx}
\usepackage[dvipsnames,svgnames,table]{xcolor}
\usepackage[figuresright]{rotating}
\usepackage{xtab} % tackle the long tables
\usepackage{longtable} % tackle the long tables
\usepackage{multirow}
\usepackage{footnote}
\usepackage[stable]{footmisc}
\usepackage{chngpage} % allows for temporary adjustment of side margins
\usepackage{pdflscape} % landscape environment
\usepackage{tocbibind} % includes everything in ToC

\usepackage{pgfplots}
\pgfplotsset{compat=1.18}
\pgfplotsset{every tick label/.append style={font=\footnotesize}}
\usepackage{setspace}
\makesavenoteenv{tabular}
\usepackage{tabularx}
\usepackage{booktabs}
\usepackage{threeparttable} % [flushleft]
\usepackage[referable]{threeparttablex} % footnotes in tabu
\newcolumntype{R}{>{\raggedleft\arraybackslash}X}
\newcolumntype{L}{>{\raggedright\arraybackslash}X}
\newcolumntype{C}{>{\centering\arraybackslash}X}
\newcolumntype{A}{>{\columncolor{gray!25}}C}
\newcolumntype{a}{>{\columncolor{gray!25}}c}

\newlength{\tablen}

\usepackage{dcolumn} % alignment to decimal points
\newcolumntype{.}{D{.}{.}{-1}}

\usepackage{tikz}
\usetikzlibrary{arrows, calc, matrix, patterns, positioning, trees}
\usepackage[semicolon]{natbib}
\usepackage[hyphens]{url}
\usepackage{hyperref} % [hidelinks]
\hypersetup{
  colorlinks   = true,    		% Colours links instead of ugly boxes
  urlcolor     = blue,    		% Colour for external hyperlinks
  linkcolor    = blue,			% Colour of internal links
  citecolor    = ForestGreen,		  	% Colour of citations
}
\usepackage[nopatch=footnote]{microtype}
\usepackage[justification=centering]{caption} % [justification=centering]

\usepackage[labelformat=simple]{subcaption}

\DeclareCaptionLabelFormat{parenthesis}{(#2)}
\makeatletter
\renewcommand\p@subfigure{\arabic{figure}.}
\makeatother

\DeclareCaptionLabelFormat{parenthesis}{(#2)}
\makeatletter
\renewcommand\p@subtable{\arabic{table}.}
\makeatother

\usepackage{alphalph}

\def\addlegendimage{\csname pgfplots@addlegendimage\endcsname}

\usepackage{enumitem}

\setlist[itemize]{leftmargin=2.5\parindent}
\setlist[enumerate]{leftmargin=2.5\parindent}

\theoremstyle{plain}
\theoremstyle{definition}

\theoremstyle{remark}

\def\keywords{\vspace{.5em} % Add keywords
{\noindent \textit{Keywords}: }}

\def\AMS{\vspace{.5em} % Add keywords
{\noindent \textbf{\emph{MSC} class}: }}

\def\JEL{\vspace{.5em} % Add keywords
{\noindent \textbf{\emph{JEL} classification number}: }}

\title{Swiss-system chess tournaments and unfairness}

\author{\href{https://sites.google.com/view/laszlocsato}{L\'aszl\'o Csat\'o}\thanks{~Corresponding author. Email: \emph{laszlo.csato@sztaki.hun-ren.hu} \newline
Institute for Computer Science and Control (SZTAKI), Hungarian Research Network (HUN-REN), Laboratory on Engineering and Management Intelligence, Research Group of Operations Research and Decision Systems, Budapest, Hungary \newline
Corvinus University of Budapest (BCE), Institute of Operations and Decision Sciences, Department of Operations Research and Actuarial Sciences, Budapest, Hungary}
$\qquad \qquad$
\href{https://sites.google.com/view/alex-krumer/}{Alex Krumer}\thanks{~Email: \emph{alex.krumer@himolde.no} \newline
Molde University College, Faculty of Business Administration and Social Sciences, Molde, Norway}}

\begin{document}
\maketitle
\thispagestyle{empty}
%\Dedication

\begin{abstract}
\noindent
The Swiss system is an increasingly popular competition format as it provides a favourable trade-off between the number of matches and ranking accuracy. However, there is no empirical study on the potential unfairness of Swiss-system chess tournaments if an odd number of rounds is played. To analyse this issue, our paper compares the number of points scored in the tournament between players who played one game more with the white pieces and players who played one game fewer with the white pieces. Using data from 28 highly prestigious competitions, we find that players with an extra white game score significantly more points. In particular, the advantage exceeds the value of a draw in the four Grand Swiss tournaments. A potential solution to this unfairness could be organising Swiss-system chess tournaments with an even number of rounds, and guaranteeing a balanced colour assignment for all players using a recently proposed pairing mechanism.

% first-mover advantage

\AMS{62F07, 62P20, 90B90, 91B14}
% Ranking and selection
% Applications of statistics to economics
% Case-oriented studies in operations research
% Social choice

\JEL{C44, D71, Z20}
% Operations Research, Statistical Decision Theory
% Social Choice, Clubs, Committees, Associations
% Sports Economics, General
\end{abstract}

\clearpage
\restoregeometry

\section{Introduction} \label{Sec1}

Ensuring fairness is a key responsibility of tournament organisers. Although fairness has several different interpretations in the literature, any sports rule can be judged unfair if it violates equal treatment of equals, and some players enjoy an advantage over the other players due to random factors. For example, penalty shootouts in football are widely accepted to be fair only if the outcome of the coin toss---that determines the shooting order---does not affect their outcome \citep{ApesteguiaPalacios-Huerta2010, KocherLenzSutter2012, Palacios-Huerta2014, BramsIsmail2018, LambersSpieksma2021, VollmerSchochBrandes2024, Pipke2025, vanHemertvanderKamp2026}. Our study aims to analyse Swiss-system chess competitions from this perspective.

The Swiss system is a non-eliminating tournament format where the contestants play a predetermined number of rounds, which is substantially smaller than required by a round-robin tournament. The Swiss system is especially popular in chess; indeed, it was invented by \emph{Dr.~Julius M\"uller}, a Swiss teacher, who developed a new pairing system in 1895 for the 5.~Schweizerische Schachturnier in Zurich \citep{Schulz2020}. It is now applied in other sports, too, such as badminton, croquet, and various esports \citep{DongRibeiroXuZamoraMaJing2023}, and has inspired the incomplete round-robin format of UEFA club competitions, introduced in the 2024/25 season \citep{Gyimesi2024, DevriesereGoossensSpieksma2026}.

The Swiss system is a reasonable choice if
(i) a high number of contestants makes a round-robin tournament infeasible due to time constraints; and
(ii) eliminating any contestant before the end of the tournament is undesirable.
According to the Monte Carlo simulations of \citet{SziklaiBiroCsato2022}, the Swiss system offers the best trade-off between the number of matches and the ability of the competition to reproduce the latent true ranking of the players, especially if all contestants should be ranked. This optimality provides a strong motivation for investigating the Swiss-system tournaments in depth. 

%Swiss-system tournaments can contain either an even or an odd number of rounds.
In chess, most Swiss-system tournaments contain an odd number of rounds, typically nine or 11. The default length is probably used because the difference between the number of black and white games should be between $-2$ and $+2$ according to the FIDE (F\'ed\'eration Internationale des \'Echecs, International Chess Federation) pairing rules. Hence, Swiss-system tournaments with an even number of rounds are undesirable because they allow for playing two more games with the black or white pieces \citep{Schulz2020}. For example, in the Gibraltar International Chess Festival 2020 - Masters tournament (\url{https://chess-results.com/tnr471965.aspx}), 5-5 players had 4 and 6 white games out of the total 10, respectively. 

However, with an odd number of rounds, (about) half of the players play more games with the white pieces than the other players.
This might be unfair since the player with the white pieces starts the game and enjoys an advantage over their opponent \citep{Fecker2024, Gonzalez-DiazPalacios-Huerta2016, Henery1992, Milvang2015}. Evidence of white advantage appears in modern computer chess, too: when AlphaZero---a computer program developed by an artificial intelligence research company to master the game of chess---played 10 thousand games against itself at 1 second per move, White won 65\% of decisive games, although 88.2\% of the games ended in a draw \citep[Figure~2a]{TomasevPaquetHassabisKramnik2020}.
%Even the rules of FIDE reflect white advantage to some extent. In the Armageddon decisive game, used as a final tiebreaker after multiple draws in the previous stages of a match, black starts with less time on the clock than white, but a draw counts as a win for black. 

Unsurprisingly, \citet{BramsSeven2025} note that 66\% of the top 10 finishers in major Swiss-system tournaments had an extra white game. Analogously, in the four editions of probably the most important Swiss-system chess tournament, the FIDE Grand Swiss, 22 players scored at least 7.5 points, and 20 of them (90.9\%) played one game more with the white pieces.

Inspired by this anecdotal evidence, our paper aims to quantify the ex post advantage provided by the extra game with the white pieces; without establishing the causal existence of white advantage itself, which is already well documented in the literature.
To the best of our knowledge, this is the first paper that makes such an attempt.
For that, data from 28 top-level Swiss-system tournaments played between 2014 and 2025 with an average Elo rating of at least 2400 are used. We find that playing more games with the white pieces significantly increases the number of points scored in the tournament. The effect is slightly above 0.25 points for all tournaments and above 0.5 points for the four Grand Swiss tournaments, when wins, draws, and losses are awarded by 1, 0.5, and 0 points, respectively. 
This robust advantage suggests a severe \emph{ex post} fairness problem of the Swiss system with its current rules in professional chess.
%assigns the white games essentially randomly. Therefore, some lucky players benefit  from a kind of random ``coin toss''.

The paper is structured as follows.
Section~\ref{Sec2} presents the main characteristics of the Swiss system, and Section~\ref{Sec3} gives a concise overview of previous research on this field. The data are described in Section~\ref{Sec4} and the methodology in Section~\ref{Sec5}. The results are reported in Section~\ref{Sec6}. Finally, Section~\ref{Sec7} summarises our message for governing bodies in sports and offers a potential solution to the unfairness of Swiss-system chess tournaments.

\section{The Swiss system in chess} \label{Sec2}

The two central design issues in Swiss-system tournaments are
(i) how to pair the contestants in each round; and
(ii) how to rank the contestants based on their previous results.

In Swiss-system chess tournaments, the pairing should satisfy three hard constraints \citep{FIDE2020}:
\begin{itemize}
\item
Two players cannot play against each other more than once;
\item
The difference between the number of black and the number of white games should be between $-2$ and $+2$ for each player;
\item
No player is allowed to receive the same colour three times in a row.
\end{itemize}
The last two constraints may exceptionally be violated in the last round of a tournament.

Furthermore, three soft criteria exist that can be used to evaluate a pairing mechanism:
\begin{itemize}
\item
In general, players are paired to others with the same score;
\item
In general, a player is given the colour with which they played fewer games;
\item
If colours are already balanced, then, in general, a player is given the colour that alternates from the last one.
\end{itemize}
Colour allocation in the first round is random for each match separately.

Currently, FIDE accepts four variants of the Swiss system, the classical Dutch system, as well as the Lim, Dubov, and Burstein systems, which differ in their pairing method \citep{FIDE2020}. The Dutch and Dubov systems are compared via simulations by \citet{Held2020}.
Pairing has traditionally been done by hand, without computer assistance; today, FIDE endorses some tournament manager software to organise Swiss-system competitions, see \url{https://spp.fide.com/endorsed-tournament-managers/}. Nevertheless, due to the complex rules, even pairing engines endorsed by FIDE make occasional mistakes \citep{SauerCsehLenzner2024}.
%For example, the Swiss-Manager (\url{https://swiss-manager.at/}) can handle competitions up to 60 players and 11 rounds for a price of 75 Euro, and up to 1,500 players (or 300 teams) and 23 rounds for 150 Euro.

In chess, every match has three possible outcomes: white wins (black loses), white loses (black wins), and draw. The winner receives 1 point, the loser 0 points, and a draw gives 0.5-0.5 points to both players. Thus, the total number of points for any player in a tournament could be between 0 and the number of rounds with a step of 0.5. The ranking is based on the total number of points scored, followed by various tie-breaking criteria \citep{Freixas2022}. Tie-breaking rules are not considered in our study.

Almost all players have an Elo rating, a measure of strength based on past performances \citep{Elo1978, Aldous2017, GomesdePinhoZancoSzczecinskiKuhnSeara2024}.
Let $R_i$ and $R_j$ denote the Elo ratings of players $i$ and $j$, respectively. The expected result of their match is 
\begin{equation*}
E_{ij} = \frac{1}{1 + 10^{\left( R_j - R_i \right) /400}}.
\end{equation*}
Note that $E_{ji} = 1 - E_{ij}$.
The actual scores $S_{ij}$ and $S_{ji} = 1 - S_{ij}$ determine the updates of the ratings $R_i' = R_i + K \left( S_{ij} - E_{ij} \right)$ and $R_j' = R_j + K \left( S_{ji} - E_{ji} \right)$. Consequently, $R_i + R_j = R_i' + R_j'$. The maximum possible adjustment per game, $K$, is set to 10 once the published rating of a player reaches 2400, and remains at that level subsequently.

In a tournament with $2m+1$ rounds, where $m$ is a positive integer, about half of the players play $m+1$ games with the white pieces and $m$ games with the black pieces, and the other half of the players play $m$ games with the white pieces and $m+1$ games with the black pieces.
On the other hand, the FIDE rules do not exclude the possibility that a player has $m+1$ or $m-1$ games with the white pieces in a tournament with $2m$ rounds. We will return to this issue in Section~\ref{Sec7}.

\section{Related literature} \label{Sec3}

The pairing methods of Swiss-system tournaments have received serious attention in the operations research literature.
\citet{Olafsson1990} provides an efficient computer program to pair the players in each round. It is based on finding a maximum weight matching in a graph, where the players are nodes and the possible matches are edges such that the pairing rules are converted into edge weights.
\citet{KujansuuLindbergMakinen1999} translate the main principles behind the pairing in the Swiss system (players with roughly equal scores play against each other and their colours alternate) into a stable roommates problem. This approach results in better colour balance but higher score differences than the official FIDE pairing.

\citet{GlickmanJensen2005} propose adaptive pairings by maximising the expected Kullback--Leibler distance between the prior and posterior densities of strength parameters. In contrast to the Swiss system that tends to pair players of similar strength in later rounds, their algorithm pairs players of similar strength as soon as possible. The pairing mechanism outperforms the modified Swiss system of \citet{Olafsson1990} if 16 rounds are played by 50 players, regardless of using a vague or an informative prior distribution. However, we do not know of any Swiss-system chess tournament containing more than 11 rounds. 
\citet{BiroFleinerPalincza2017} uncover how the main priority rules of the Dutch system can be replaced by efficient matching algorithms, making computation in polynomial time possible.

In contrast to \citet{Olafsson1990}, \citet{SauerCsehLenzner2024} formulate alternative pairing rules, again enforced by finding maximum weight matchings in an appropriate graph. According to extensive numerical simulations, this procedure yields fairer pairings and leads to a final ranking that better reflects the strengths of the players than the official FIDE pairing system. Furthermore, the proposed method allows for great flexibility by changing edge weights. In particular, maximum colour difference can be reduced by allowing more games between players who have scored different numbers of points \citep[Section~4.4]{SauerCsehLenzner2024}. This trade-off will be important for us in Section~\ref{Sec7}.

A unique feature of the Swiss system, dynamic scheduling, can also be advantageous in other sports.
For instance, \citet{DongRibeiroXuZamoraMaJing2023} develop and evaluate dynamic scheduling methods for e-sports tournaments based on the Swiss system design: an integer programming model is used to maximise game attractiveness and the utility of spectators.
\citet{DiMattiaKrumer2026} demonstrate that reducing the number of matches per team in groups of four teams from three (round-robin) to two (Swiss system) in beach volleyball does affect neither the probability of the favourite team winning a single match, nor the overall efficacy of the tournament.

Three empirical studies in chess are also closely related to our topic.
The first is \citet{LinnemerVisser2016}, who find that, when lowly rated players play against highly rated players in a Swiss-system tournament, the former perform better than predicted by the Elo formula (see Section~\ref{Sec2}). The reason is self-selection: the expected strength-shock, which is private information, decreases with the Elo rating because lowly rated players enter the tournament only if they observe their actual strength to be higher than their Elo rating.

The second paper, \citet{Gonzalez-DiazPalacios-Huerta2016} analyse 197 chess matches with an even number of games between two opponents, where the colour of the pieces alternates in each round. Even though there is no rational reason why winning frequencies should differ from 50-50\%, winning probabilities turn out to be about 60-40\%, favouring the player who plays with the white pieces in the first and subsequent odd rounds.

Interestingly, FIDE has moved in the opposite direction as proposed by \citet{Gonzalez-DiazPalacios-Huerta2016}. From 2008 to 2018, the World Chess Championships were contested by the reigning champion and the challenger, such that one player played with the white pieces in the odd games in the first part of the match, while this pattern was reversed in the second half. However, since the World Chess Championship 2021, the player who gets white in the first game has white in all odd games.

%The recent bachelor's thesis
Finally, \citet{Fecker2024} examines more than 12.5 million chess games to explore white advantage. The results can be summarised as follows:
(1) white advantage increases as a function of Elo rating (at the level of Grandmasters, it is approximately twice as high as at the level of amateurs);
(2) white advantage has generally remained stable, or even slightly increased, in all categories from 2010 to 2023;
(3) the only exception is for games with an average Elo rating above 2600, where white advantage has decreased over time.

%Last but not least, the current paper contributes to the rapidly growing literature on fairness in tournament design. Recent surveys \citep{KendallLenten2017, GoossensYiVanBulck2020, LentenKendall2022, DevriesereCsatoGoossens2025, DevriesereGoossensWillem2025} and a book \citep{Csato2021a} provide an insight into this area.

\section{Data and variables} \label{Sec4}

Data are collected from top-levels individual chess tournaments between the years 2014 and 2025 that used the Swiss system with standard time control.
We have considered the following six series: Aeroflot Open, Dubai Open, (FIDE) Grand Prix, (FIDE) Grand Swiss, Qatar Masters, and Sharjah Masters. A tournament was included if it had an average Elo rating of the participants of 2400 or above, which represents the level of International Master.

Overall, our dataset consists of 2602 entries (i.e., players per tournament). However, in some cases, the players did not play in all rounds. This could happen when a player had a ``bye'', which occurs in tournaments with an odd number of players. In this case, the player currently ranked last receives a so-called half-point ``bye'' in each round---but no player can receive a ``bye'' more than once. Another situation where players did not compete in all rounds is when they simply retired before the end of the tournament; for example, because of health or travel issues. In total, we had 230 players per tournament that did not compete in all the matches because of a ``bye'' or retirement. Removing these cases, we are left with 2372 entries.

\begin{table}[t!]
  \centering
  \caption{List of the most prestigious Swiss-system chess tournaments, 2014--2025}
  \label{Table1}
    \rowcolors{1}{}{gray!20}
\begin{threeparttable}
    \begin{tabularx}{\textwidth}{l ccC CCC} \toprule
    Tournament & Average & Std.~Dev. & Min & Max & Players & Rounds \\ \bottomrule
    \href{https://chess-results.com/tnr164197.aspx}{Aeroflot 2015} & 2591  & 78    & 2379  & 2756  & 63    & 9 \\
    \href{https://chess-results.com/tnr205426.aspx}{Aeroflot 2016} & 2589  & 77    & 2395  & 2735  & 80    & 9 \\
    \href{https://chess-results.com/tnr266418.aspx}{Aeroflot 2017} & 2584  & 77    & 2418  & 2738  & 94    & 9 \\
    \href{https://chess-results.com/tnr322070.aspx}{Aeroflot 2018} & 2570  & 75    & 2386  & 2724  & 86    & 9 \\
    \href{https://chess-results.com/tnr403175.aspx}{Aeroflot 2019} & 2573  & 76    & 2426  & 2733  & 93    & 9 \\
    \href{https://chess-results.com/tnr508930.aspx}{Aeroflot 2020} & 2556  & 79    & 2402  & 2728  & 79    & 9 \\
    \href{https://chess-results.com/tnr900901.aspx}{Aeroflot 2024} & 2439  & 102   & 2270  & 2732  & 132   & 9 \\
    \href{https://chess-results.com/tnr1118221.aspx}{Aeroflot 2025} & 2437  & 120   & 2195  & 2753  & 120   & 9 \\ \hline
    \href{https://chess-results.com/tnr759378.aspx}{Dubai 2023} & 2486  & 120   & 2227  & 2729  & 63    & 9 \\
    \href{https://chess-results.com/tnr947196.aspx}{Dubai 2024} & 2455  & 92    & 2303  & 2654  & 44    & 9 \\
    \href{https://chess-results.com/tnr1155334.aspx}{Dubai 2025} & 2475  & 82    & 2277  & 2693  & 60    & 9 \\ \hline
    \href{https://chess-results.com/tnr288645.aspx}{Grand Prix 2017 Geneva} & 2728  & 46    & 2638  & 2809  & 18    & 9 \\
    \href{https://chess-results.com/tnr280762.aspx}{Grand Prix 2017 Moscow} & 2730  & 52    & 2621  & 2795  & 18    & 9 \\
    \href{https://chess-results.com/tnr307271.aspx}{Grand Prix 2017 Palma} & 2728  & 48    & 2629  & 2801  & 18    & 9 \\
    \href{https://chess-results.com/tnr263691.aspx}{Grand Prix 2017 Sharjah} & 2724  & 50    & 2628  & 2796  & 18    & 9 \\ \hline
    \href{https://chess-results.com/tnr478041.aspx}{Grand Swiss 2019} & 2618  & 101   & 2300  & 2876  & 145   & 11 \\
    \href{https://chess-results.com/tnr587230.aspx}{Grand Swiss 2021} & 2639  & 56    & 2467  & 2800  & 108   & 11 \\
    \href{https://chess-results.com/tnr793016.aspx}{Grand Swiss 2023} & 2638  & 85    & 2225  & 2786  & 111   & 11 \\
    \href{https://chess-results.com/tnr1246285.aspx}{Grand Swiss 2025} & 2641  & 69    & 2287  & 2785  & 114   & 11 \\ \hline
    \href{https://chess-results.com/tnr153113.aspx}{Qatar 2014} & 2514  & 140   & 2210  & 2776  & 152   & 9 \\
    \href{https://chess-results.com/tnr199261.aspx}{Qatar 2015} & 2533  & 136   & 2260  & 2834  & 126   & 9 \\
    \href{https://chess-results.com/tnr831193.aspx}{Qatar 2023} & 2465  & 123   & 2250  & 2839  & 148   & 9 \\
    \href{https://chess-results.com/tnr1080336.aspx}{Qatar 2024} & 2439  & 132   & 2205  & 2801  & 129   & 9 \\ \hline
    \href{https://chess-results.com/tnr579252.aspx}{Sharjah 2021} & 2533  & 135   & 2173  & 2704  & 56    & 9 \\
    \href{https://chess-results.com/tnr617267.aspx}{Sharjah 2022} & 2566  & 93    & 2365  & 2701  & 75    & 9 \\
    \href{https://chess-results.com/tnr724974.aspx}{Sharjah 2023} & 2617  & 55    & 2519  & 2734  & 74    & 9 \\
    \href{https://chess-results.com/tnr902564.aspx}{Sharjah 2024} & 2598  & 73    & 2463  & 2761  & 81    & 9 \\
    \href{https://chess-results.com/tnr1100164.aspx}{Sharjah 2025} & 2559  & 66    & 2453  & 2771  & 67    & 9 \\ \bottomrule
    \end{tabularx}
\begin{tablenotes} \footnotesize
\item
Columns Average, Std.~Dev., Min, Max show the descriptive statistics of the Elo ratings of all players.
\item
\emph{Source:} \url{https://chess-results.com/}; the hyperlink of a tournament directly navigates to the relevant URL.
\end{tablenotes}
\end{threeparttable}

\end{table}

Table~\ref{Table1} presents the 28 tournaments used in our analysis with the 2372 player-tournament observations, representing players who competed in all rounds of the respected tournaments. There were 11 rounds in the four Grand Swiss, whereas nine rounds were used in the other 24 tournaments. The average Elo rating was above 2500 in 21 tournaments (and above 2700 in the four Grand Prix tournaments), which corresponds to the level of Grandmaster, the highest title awarded by FIDE. The average Elo in the other seven tournaments is between 2400 and 2500.

\begin{table}[t!]
  \centering
  \caption{Descriptive statistics of the variables used}
  \label{Table2}
    %\rowcolors{1}{gray!20}{}
    \begin{tabularx}{\textwidth}{l cCC cCC} \toprule \hiderowcolors
    \multirow{2}[1]{*}{Variable} & \multicolumn{3}{c}{One game more with white} & \multicolumn{3}{c}{One game fewer with white} \\
    & Mean (sd) & Min & Max & Mean (sd) & Min & Max \\ \midrule
    \multirow{2}[1]{*}{Points won} & 5.033 & \multirow{2}{*}{2} & \multirow{2}{*}{8.5} & 4.571 & \multirow{2}{*}{0.5} & \multirow{2}{*}{8} \\
    & (1.041) &       &       & (1.132) &       &  \\
    \multirow{2}[0]{*}{Elo rating} & 2564  & \multirow{2}{*}{2195} & \multirow{2}{*}{2876} & 2535  & \multirow{2}{*}{2173} & \multirow{2}{*}{2812} \\
    & (122) &       &       & (126) &       &  \\
    \multirow{2}[0]{*}{First game with white} & 0.794 & \multirow{2}{*}{0} & \multirow{2}{*}{1} & 0.203 & \multirow{2}{*}{0} & \multirow{2}{*}{1} \\
    & (0.405) &       &       & (0.402) &       &  \\ \midrule \showrowcolors
    Observations & \multicolumn{3}{c}{1,198} & \multicolumn{3}{c}{1,174} \\ \bottomrule
    \end{tabularx}
\end{table}

The performance of the players in a tournament is quantified by their total number of points, which is the first criterion in the final tournament ranking (see Section~\ref{Sec2}).
Pre-tournament Elo ratings serve as the measure of abilities. In addition, we have information about the number of games played with the white pieces, and whether the first match was played with the white pieces. Table~\ref{Table2} presents the descriptive statistics of these variables, separately for the two sets of players. Players who had one game more with the white pieces scored a higher number of points on average. However, they also have a higher Elo rating and started the tournament substantially more often with the white pieces compared to players who played one game fewer with the white pieces, even though colour allocation in the first round is random according to the rules of the Swiss system.

\section{Estimation strategy} \label{Sec5}

We are interested in studying the impact of playing more games with the white pieces on the overall performance in Swiss-system chess tournaments. If the allocation of the players between the schedules with more or fewer games with the white pieces was entirely random, then we would just compare the means of the final points obtained by the players from these two groups. The difference would be a consistent estimate of the desired effect. However, such a simple approach would yield biased and inconsistent estimates in our case. The reason is that, as observed in Table~\ref{Table2}, the allocation of players who play more games with the white pieces is not determined at random. Rather, there is a selection into playing more games with the white pieces that is driven by the players' Elo ratings and whether they played the first match with the white pieces. Therefore, it is essential to disentangle the effect coming with the selection (e.g., the abilities of the players) from the effect caused by playing more games with the white pieces.

To address this issue, we use a statistical matching approach. More specifically, we apply the radius-matching-on-the-propensity score estimator with bias adjustment \citep{LechnerMiquelWunsch2011}. It is not only competitive among a range of propensity score related estimators, but the subsequent paper \citet{HuberLechnerWunsch2013} actually showed its superior finite sample and robustness properties in a large-scale empirical Monte Carlo study.

The main idea behind this estimator is to compare treated (players with more games with the white pieces) and non-treated (players with more games with the black pieces) observations within a specific radius. The first step consists of a distance-weighted radius matching on a propensity score estimation. Simply stated, we estimate a logit regression where the dependent variable is our treatment (a dummy variable that equals one if a player played more games with the white pieces and zero otherwise) and independent variables are the observable characteristics. From that estimation, we derive propensity scores (i.e., probability for each observation to be treated, given its characteristics). Then, treated and non-treated observations are matched within a predefined radius, assigning weight for each comparison based on the distance between them.
In contrast to standard matching algorithms, where controls within the radius have the same weight independent of their location, in the radius matching method, controls within the radius are weighted proportionally to the inverse of their distance to the respective treated observations to which they are matched. These weights obtained from the matching are used in a weighted linear regression in order to remove biases due to mismatches. This approach uses all comparison observations within a predefined distance around the propensity score, which allows for greater precision than the fixed nearest neighbour matching in regions where many similar comparison observations are available.%\footnote{~A number of previous studies have applied this estimator to investigate scheduling and fairness issues in sports, see, for example, \citet{Cohen-ZadaKrumerShapir2018, GollerKrumer2020, KrumerLechner2018, Krumer2020a}.} 

The analysis was executed using the \texttt{radiusmatch} command in Stata 16 \citep{HuberLechnerSteinmayr2012}. See \citet{HuberLechnerSteinmayr2015} for its details and implementation in different software packages such as Gauss, Stata, and R.

Crucially, since assignment to an extra white game is partially path dependent, our estimates should be interpreted as adjusted ex post tournament-level differences associated with receiving one additional white game, rather than as purely randomised causal effects.

\section{Empirical evidence of unfairness} \label{Sec6}

\begin{table}[t!]
  \centering
  \caption{Propensity score estimation}
  \label{Table3}
    %\rowcolors{1}{}{gray!20}
\begin{threeparttable}
    \begin{tabularx}{0.6\textwidth}{l C} \toprule
    Variable & Effect \\ \midrule
    \multirow{2}[1]{*}{First game with the white pieces} & 0.445*** \\
    & (0.004) \\
    \multirow{2}[1]{*}{Elo rating} & 0.0007*** \\
    & (0.0001) \\
	Tournaments dummies & Yes \\ \midrule
    Number of observations & 2372 \\ \bottomrule
    \end{tabularx}
\begin{tablenotes} \footnotesize
\item
Logit average marginal effects from estimating all the data. Dependent variable is a dummy of whether a player played one game more with the white pieces.
\item
Standard errors, given in parentheses, are clustered at the tournament level. Level of significance: * $p < 10\%$; ** $p < 5\%$; *** $p < 1\%$.
\end{tablenotes}
\end{threeparttable}
\end{table}

Our study aims to quantify the effect of playing more games with the white pieces on overall performance, as measured by the total number of points obtained at the end of the tournament. Table~\ref{Table3} reports the results of the propensity score estimation. Although this estimation serves only technical purposes, namely, to allow the easy purging of the results from the selection effects, it is nevertheless interesting to see which variables drive selection. In line with the descriptive results presented in Table~\ref{Table2}, Elo ratings and playing the first game with the white pieces are significantly associated with playing more games with the white pieces. In particular, players who play the first game with the white pieces are found to have a 44.5\% higher probability to play more games with the white pieces in the tournament. In addition, any additional Elo point increases the probability of playing more games with the white pieces by 0.07\%.
Tournament dummies are used in this exercise to control for the invariant features that are assumed to apply to all players (e.g., monetary prize, number of rounds, number of players, location, weather, etc.).

\begin{table}[t!]
  \centering
  \caption{Levels and effects of playing one game more with the white pieces}
  \label{Table4}
  
\begin{subtable}{\textwidth}
  \caption{All tournaments}
  \label{Table4a}
    \rowcolors{1}{}{gray!20}
    \begin{tabularx}{\textwidth}{l CCc ccc} \toprule
    Sample & Exp.~points with more white & Exp.~points with fewer white & Effect & Std.~Error & Obs. & On-supp. \\ \bottomrule
    (1) All data & 4.999 & 4.648 & 0.351*** & 0.055 & 2372 & 2362 \\
    (2) Nine Rounds & 4.776 & 4.509 & 0.267*** & 0.055 & 1894 & 1887 \\
    (3) 11 Rounds & 5.809 & 5.242 & 0.568*** & 0.107 & 478   & 455 \\ \toprule
    \end{tabularx}
\end{subtable}

\vspace{0.5cm}
\begin{subtable}{\textwidth}
  \caption{Tournaments with average Elo above 2500}
  \label{Table4b}
    \rowcolors{1}{}{gray!20}
\begin{threeparttable}
    \begin{tabularx}{\textwidth}{l CCc Ccc} \toprule
    Sample & Exp.~points with more white & Exp.~points with fewer white & Effect & Std.~Error & Obs. & On-supp. \\ \bottomrule
    (4) All data & 5.041 & 4.681 & 0.359*** & 0.061 & 1676 & 1670 \\
    (5) Nine Rounds & 4.707 & 4.476 & 0.231*** & 0.062 & 1198 & 1195 \\ \bottomrule
    \end{tabularx}
\begin{tablenotes} \footnotesize
\item
Average treatment effect estimated using radius matching on the propensity score. 
\item
Obs.: Total sample used in the estimation. On-supp.: Effective sample after removing units that are off common support.
\item
Standard errors are clustered at the tournament level. Level of significance: * $p < 10\%$; ** $p < 5\%$; *** $p < 1\%$.
\end{tablenotes}
\end{threeparttable}
\end{subtable}

\end{table}

Table~\ref{Table4} shows the results of the radius matching average impact of playing one game more with the white pieces on the number of points scored at the end of the tournament. Each line represents the estimations on different subsamples. Based on all the data (Line 1), players with one game more with the white pieces achieve 0.351 more points on average than players with one game less. This estimate is significant at the 1\% level.
In tournaments with nine rounds (Line 2), players with one game more with the white pieces score on average 0.267 points more than their counterparts with one game less. However, the size of this effect in tournaments with 11 rounds is more than doubled to 0.568 points (Line 3).
Both are significant at the 1\% level.

Finally, we present the estimations for tournaments where the mean Elo rating is above 2500 points, which represents a level of Grandmaster. Seven tournaments with a mean Elo rating between 2400 and 2500 are excluded, which leaves us with 21 tournaments (see Table~\ref{Table1}). The results for this subset, reported in Table~\ref{Table4b}, are analogous to the results for all tournaments in Table~\ref{Table4a}.\footnote{~Note that all tournaments with 11 rounds have an average Elo rating above 2600.}

\section{Discussion} \label{Sec7}

We have examined Swiss-system chess tournaments with an odd number of rounds from the perspective of ex post fairness. These competitions turn out to be unfair as players who play more games with the white pieces enjoy a substantial advantage.
%Our empirical findings can be relevant to all sports where home advantage---which is equivalent to playing with the white pieces in chess---exists, and some contestants benefit from home advantage more than the others. For example, the new incomplete round-robin format of UEFA club competitions will certainly lead to great controversies among the fans of the clubs if it would consist of seven rounds instead of the current six or eight.

At first sight, a simple solution could be organising Swiss-system tournaments with an even number of rounds. Nevertheless, as we have discussed in the introduction, the current FIDE rules cannot guarantee a perfect colour balance for these tournaments. Consequently, if even number of rounds would be used without any change in the pairing mechanism, the maximal level of unfairness certainly worsens because some players might have two more games with the white pieces.

An attractive solution to this problem is provided by the mechanism of \citet{SauerCsehLenzner2024}, which computes maximum weight matchings in an auxiliary weighted graph, where two criteria, pairing players within their score groups and colour balance, have associated weights. Due to this flexibility, a stronger colour difference constraint can be easily enforced by changing the weights towards colour balance. Even an alternating black-white sequence can be guaranteed for all players with an appropriate setting.
According to the simulations, this version achieves the same ranking quality (the similarity of the final ranking to the ranking of the players by their Elo ratings) as the official FIDE pairing mechanisms \citep{SauerCsehLenzner2024}.

Obviously, enforcing colour balance comes at a price: the number of floaters, players who play against a player having a different score, is slightly increased. While this might be considered to be against the logic of the Swiss system, it is important to note that the number of scores is \emph{not} the best measure of performance \citep{Csato2017a}. Hence, a higher number of floaters does not seem to be a serious problem if the ranking quality remains unchanged.

\citet{SauerCsehLenzner2024} conclude that tournament organisers face a choice between reducing the number of floaters with the standard colour difference or an alternating colour assignment for all players with a higher number of floaters. Crucially, the ranking quality is not affected by this trade-off. The current policy of FIDE favours a lower number of floaters.
However, based on our findings of such a robust advantage provided by an extra game with the white pieces and the slight increase of the number of floaters without any change in the ranking quality \citep{SauerCsehLenzner2024}, it is plausible to expect that perfect colour balance is desirable since the gain in fairness from equalising colours outweighs the downsides of more float games. Thus, it is worthwhile considering an even number of rounds when organising Swiss-system chess tournaments.

But there can be another argument against a perfect colour balance with an even number of rounds. Guaranteeing alternating colour assignment for all players implies that the set of matches will be a bipartite graph. Consequently, if the top $k$ players have the white or black pieces in the first round, they will never play against each other during the tournament. Naturally, such a tournament may become unattractive.

On the other hand, the lack of matches between the top $k$ players can be easily mitigated by changing the policy of random colour assignment in the first round. For example, it can be required that the highest-ranked player plays with the white pieces, the second and the third play with the black pieces, the fourth and the fifth play with the white pieces, and so on. An analogous rule can effectively decrease the chance of clustering players from one of the two sets around any position in the final ranking.

Another potential solution is offered by the Leagues Cup, an international football tournament contested by US and Mexican clubs. Since 2025, its first stage is an incomplete round-robin tournament, where each of the 18 clubs from the United States and 18 clubs from Mexico play three games against three different teams from the other country. Therefore, the set of matches is a bipartite graph. Then, the clubs from the two nations are ranked separately, and the top four teams in each country advance to the knockout stage.
Swiss system chess tournaments can be organised analogously, by producing two rankings, separately for the players who start with the black and the white pieces. In order to have an overall winner, the top two players of both rankings can go to the semifinals, or the top four to the quarterfinals.

To summarise, our empirical results show an adjusted ex post tournament-point advantage associated with receiving one additional white game in elite Swiss-system chess tournaments. This finding calls for changing the design of these tournaments, and can inspire future research on the pairing rules in the Swiss system.

\section*{Acknowledgements}
\addcontentsline{toc}{section}{Acknowledgements}

We are grateful to \emph{Mehmet Mars Seven}, who have mentioned this potential problem of the Swiss system during the 3rd Fairness in Sports workshop at Ghent. \\
Two reviewers, \emph{Kjetil K.~Haugen} and three anonymous colleagues have given useful remarks. \\
The research was supported by the National Research, Development and Innovation Office under Grants Advanced 152220 and FK 145838, and the J\'anos Bolyai Research Scholarship of the Hungarian Academy of Sciences.

\bibliographystyle{apalike} 
\bibliography{All_references}

\end{document}